\documentclass[
reprint,
amsmath,amssymb,aps,prb,floatfix
]{revtex4-2}

\usepackage{physics}
\usepackage{float}
\usepackage[dvipsnames]{xcolor}
\usepackage{graphicx}
\usepackage{bm}
\usepackage{hyperref}
\hypersetup{colorlinks=true,allcolors=RoyalBlue}
\allowdisplaybreaks

\begin{document}
\title{Local heat current flow in the ballistic phonon transport of graphene nanoribbons}
\author{Yin-Jie Chen}
\affiliation{School of Physics, Institute for Quantum Science and Engineering and Wuhan National High Magnetic Field Center,
Huazhong University of Science and Technology, Wuhan 430074,  China}
\author{Jing-Tao L\"u}
\email{jtlu@hust.edu.cn}
\affiliation{School of Physics, Institute for Quantum Science and Engineering and Wuhan National High Magnetic Field Center,
Huazhong University of Science and Technology, Wuhan 430074,  China}

\begin{abstract}
Utilizing the non-equilibrium Green’s function method, we study the local heat current flow of phonons in nanoscale ballistic graphene nanoribbon, where boundary scattering leads to the formation of atomic-scale current vortices. We further map out the atomic temperature distribution in the ribbon with B\"uttiker's probe approach. From the heat current and temperature distribution, we observe inverted temperature response in the ribbon, where the heat current direction goes from the colder to the hotter region. Moreover, we show that atomic scale defect can generate heat vortex at certain frequency, but it is averaged out when including contributions from all the phonon modes. Meanwhile, our results have recovered residual-resistivity dipole features manifested at the vicinity of local defects. These results extend the study of local heat vortex and negative temperature response in the bulk hydrodynamic regime to the atomic-scale ballistic regime, further confirming boundary scattering is crucial to generate backflow of heat current.
\end{abstract}

\maketitle

\section{Introduction}

Quantum transport properties of nanoscale systems have received intense attention in the past several decades. 
Single molecular junction is one example of such systems, enabling the study of fundamental physical phenomena at atomic level, especially the out-of-equilibrium transport properties of heat and charge \cite{xin2019,dubi2011,segal2016,evers2020}. For electronic transport, while most of the work focused on global transport properties, some efforts were made to unveil local contributions of electron charge current within molecular junctions \cite{ernzerhof2006,solomon2010,rai2010,stuyver2017,nozaki2017,rix2019,pohl2019,jensen2019,stegmann2020}. The investigation of local current provides insight into the detailed behavior of electrons in the molecular junction, such as quantum interference and vortex dynamics \cite{rai2010,nozaki2017,rix2019,stegmann2020}. 

In larger two-dimensional (2D) atomic structures, local currents are calculated to achieve quantitative understanding of transport mechanisms and support the design of effective nanoscale devices \cite{todorov1999,nikolic2006,zarbo2007,liu2008,jiang2009,chen2010,kumar2010,chang2012,walz2014,wilhelm2014a,dang2015,walz2015,he2016,stegmann2016,morr2017,wang2019,zhang2019a,shao2020,yang2020a,gomes2021,shao2021,sanchez2022}.
Recently, electron current vortices have been intensively explored as a signature of hydrodynamic transport in 2D systems \cite{bandurin2016,aharon2022,palm2024}, which induces negative non-local electrical resistance \cite{wang2016,levitov2016,pellegrino2016,chandra2019,danz2020,gupta2021}. Actually, these novel transport behaviors were predicted to appear in ballistic transport regime a long time ago \cite{todorov1999}, and have been discussed again recently in comparison to hydrodynamic transport \cite{wang2016,wang2019,gomes2021}.

On the other hand, thermal transport in 2D materials has also received intense research interest \cite{gu2018,libw2022}. Theoretical works demonstrated that phonon heat vortices can exist in both hydrodynamic and ballistic regimes \cite{Cepellotti2015,Lee-2015,guo2021,zhang2021,raya2022}. Motivated by the intriguing thermal features provided in previous works, we extend these ideas to nanoscale atomic structures, where wave property of phonons becomes important. We study the local heat current and temperature distribution in graphene nanoribbons, paying special attention to the effects of boundary scattering and atomic defect \cite{hage2020a,xu2023}. 

\begin{figure}[t]
    \centering
    \includegraphics[scale=0.73]{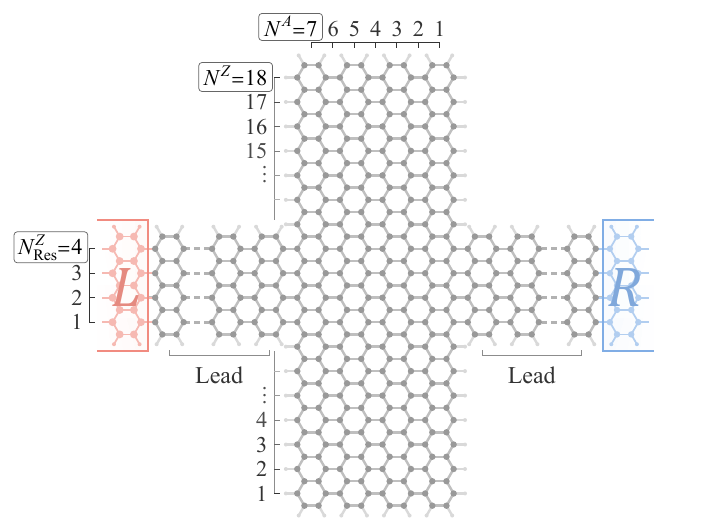}
    \caption{A sketch of the two terminal graphene nanoribbon. The reservoirs ($L$ and $R$) are connected to the ribbon through narrower leads, such that heat can be locally injected or extracted from the ribbon. $N^Z$, $N^Z_{\rm Res}$ and $N^A$ are geometric parameters.}
    \label{fig:sketch}
\end{figure}

We use the non-equilibrium Green's function (NEGF) method to describe coherent phonon transport \cite{wang2008}.
Specifically, we derive the expression for the local heat current, calculate heat flow in real space and further study the emerging heat vortices. 
Different from electronic transport where the current is mainly contributed by electrons near the Fermi level, the phonon heat current includes contribution from all the phonon modes. Thus, the emergence of heat vortices is not obvious. We analyze how the vortices depend on the geometric factors of the ribbon, the way heat is injected and extracted from the system, and how they are influenced by atomic disorder. Then, by calculating the local heat current in the ribbon with defects, we compare the results obtained by earlier studies that suggested the formation of vortices by specific local defect states \cite{yamamoto2008,bao2023}.

In addition to the heat flow, we also study the local temperature distribution, which is of great interest in non-equilibrium thermodynamics. 
The local temperature can be defined in various ways \cite{ventra2019}. In classical simulations, it is defined using the kinetic energy of atomic degrees of freedom from the equipartition theorem \cite{dhar2008,kannan2012}. Analogous to thermometers, the B\"uttiker probe and self-consistent reservoir have been used to determine local temperature when they reach local equilibrium with the system of interest \cite{buttiker1985,buttiker1986,lu2007,dubi2009a,dubi2009b,bandyopadhyay2011,saaskilahti2013,miao2016,stafford2016a,stafford2016b,behera2021}. The statistical definition for temperature \cite{braga2005,ventra2009} and fluctuation-dissipation theorem \cite{caso2010,caso2012,stafford2014} have also been generalized to express local temperature. Although these definitions give consistent results at equilibrium, they do not necessarily coincide with each other in non-equilibrium systems. In terms of experiments, local temperature can be measured in tens-of-nanometer-scale resolution using scanning thermal microscopy technique \cite{zhang2020}. This setup is similar to the idea of the B\"uttiker probe, which we use here \cite{stafford2016b}.

\section{Theory}
\subsection{Local heat current in harmonic system} \label{sec:localheatcurrent}

We start with a Hamiltonian $H$ within the harmonic approximation:

\begin{equation}
    H=\sum_{n,i} \left(\frac{1}{2}  \dot{u}_{ni}^2 +\sum_{m,j} \frac{1}{2}u_{ni} K_{ni,mj} u_{mj}\right).
    \label{H_total}
\end{equation}

Here we use the mass-normalized displacements $u_{ni}=\sqrt{m_n} x_{ni}$, $x_{ni}$ is the displacement away from the equilibrium position of $n$-th atom along the $i$ axis ($i,j=x,y,z$) and the indices run over all degrees of freedom in the system. The spring constant matrix has the symmetry $K_{ni,mj} = K_{mj,ni}$. In the quantum regime, $u_i$ and $\dot{u}_i$ turn into operators, satisfying the commutation relation $[u_{ni},\dot{u}_{mj}]=i\hbar\delta_{nm}\delta_{ij}$. 
The system is divided into three parts: the central region with finite degrees of freedom driven out of equilibrium, and the left ($L$) and right ($R$) reservoirs with infinite degrees of freedom in equilibrium. 

We then rewrite the total Hamiltonian as: 

\begin{align}
H=\sum_{\alpha=L, R} (H_\alpha+H_{\alpha C})+H_C,
\end{align}
with
\begin{align}
    H_\alpha&=\sum_{i,n\in\alpha} (\frac{1}{2}  \dot{u}_{ni}^2+\sum_{j,m\in \alpha} \frac{1}{2} u_{ni }K_{ni,mj}^\alpha  u_{mj}),
    \\
    H_C&=\sum_{i,n\in C}( \frac{1}{2}  \dot{u}_{ni}^2+\sum_{j,m \in C} \frac{1}{2} u_{ni} K_{ni,mj}^C u_{mj}),
    \\
    H_{\alpha C}&=\sum_{\substack{i,n\in C\\ j,m\in\alpha}}\frac{1}{2} (u_{ni} V_{ni,mj} u_{mj} +  u_{mj} V_{mj,ni} u_{ni}),
\end{align}
where $C$ denotes the central region, and $V_{ni,mj}=V_{mj,ni}$ is the coupling matrix of the reservoir and the central region. 

Local heat current $J_{nm}$ from atom site $n$ to site $m$ in the central region can be defined intuitively from the continuity equation 

\begin{equation}
    \frac{\partial h_n}{\partial t} +\sum_{m(\neq n)} J_{nm}=0,
\end{equation} 

where $h_n$ is the local energy. In this work, we choose $h_n$ containing kinetic energy terms of atom $n$, all coupling terms with reservoirs, and half of the harmonic terms between atom sites:

\begin{align}
\label{label1}
h_n &= \sum_i \frac{1}{2} \Big[  \dot{u}_{ni}^2 + \sum_{j,m\in C} u_{ni} K_{ni,mj}^C u_{mj} \notag
\\
& \quad+ \sum_{j,m\in \alpha} (u_{ni} V_{ni,mj} u_{mj} + u_{mj} V_{mj,ni} u_{ni}) \Big]. 
\end{align} 

Now, the decomposition of the total Hamiltonian becomes $H=\sum_\alpha H_\alpha + \sum_{n \in C} h_n $, similar definitions were used in other works \cite{lepri2003,segal2008,saaskilahti2012,tuovinen2016,chien2019,michelini2019,nitzan2020-1,nitzan2020-2,segal2020,karaslimane2020,chien2022}. Then, from the Heisenberg equation of motion, we get

\begin{align}
\label{label2}
-\frac{\partial h_n}{\partial t} &= \frac{i}{\hbar}[h_n , H]\notag
\\
&= \frac{1}{2} \sum_{i,j,m\in C}(u_{mj}  K_{mj,ni}^C \dot{u}_{ni} - u_{ni}  K_{ni,mj}^C \dot{u}_{mj}) \notag
\\ 
& \quad- \frac{1}{2} \sum_{i,j,m\in \alpha}  (u_{ni} V_{ni,mj} \dot{u}_{mj} + \dot{u}_{mj} V_{mj,ni} u_{ni}).
\end{align}

Here, the first term after the second equal sign represents the local heat current flowing out of site $n$, the second term equals to the heat current $J_\alpha$ from reservoir $\alpha$ in steady state. 
The definition of local energy in Eq. (\ref{label1}) is not unique \cite{campbell2000,martin-moreno2007,hardy1963,mathews1974}, however, the degrees of freedom selected for $h_n$ are simply different views (or resolutions) to study the nanosystem, having no effect on the energy flow in the harmonic quantum network since phonon dynamics is established as the Hamiltonian is defined. Our present choice is the simplest scheme and ensures the consistency with the expression used in the standard NEGF method, where the very same expression for phonon heat current is obtained for the heat current injected into the phonon bath $J_\alpha=\langle -\partial H_\alpha/\partial t\rangle$ \cite{yamamoto2006,wang2008,wang2014,wang2006,wang2007}.
Moreover, the expression for local heat current here is in analogy to the electrical current expression in electron transport, where partition of the hopping term is not needed. Rewriting the formula for local heat current in the form of Green's functions \cite{mingo2003,mingo2006,yamamoto2008,kam2009,yang2011,luisier2012,duan2018,zhou2018,luisier2018,luisier2020}, we have
\begin{align}
\label{localheatcurrent1}
    J_{nm}=\frac{i\hbar}{2}\frac{\partial}{\partial t}&\sum_{i,j}\Big[K_{mj,ni}^C G^<_{ni,mj}(t,t') \notag
    \\
    &\qquad -K_{ni,mj}^C G^<_{mj,ni}(t,t') \Big] |_{t'\to t},
\end{align}
where $i\hbar G^<_{ni,mj}(t,t')=\langle u_{mj}(t') u_{ni}(t)\rangle$ is the lesser Green's function, and the greater Green's function $i\hbar G^>_{ni,mj}(t,t')=\langle u_{ni}(t) u_{mj}(t')\rangle$ can be defined accordingly. In steady state, it is convenient to work in the frequency domain, the lesser Green's function satisfies the following equation \cite{jauho}, which is expressed in terms of the retarded Green's function $\boldsymbol{G}^r(\omega)$, the advanced Green's function  $\boldsymbol{G}^a(\omega)$, and the lesser self-energy $\boldsymbol{\Sigma}^<(\omega)$ in matrix form: 
\begin{align}
\label{greensfunc1}
\boldsymbol{G}^<(\omega)&=\boldsymbol{G}^r(\omega)\boldsymbol{\Sigma}^<(\omega)\boldsymbol{G}^a(\omega),\\
\boldsymbol{G}^r(\omega) &=[(\omega+i\eta)^2-\boldsymbol{K}^C-\boldsymbol{\Sigma}^r(\omega)]^{-1},\\
\boldsymbol{G}^a(\omega) &=[\boldsymbol{G}^r(\omega)]^\dagger.
\end{align}
Here, $\boldsymbol{\Sigma}^{<,r,a}(\omega)= \sum_\alpha \boldsymbol{\Sigma}^{<,r,a}_\alpha(\omega)$ are the lesser, retarded and advanced self-energies due to the coupling to the reservoirs. 
Introducing the Bose-Einstein distribution function $f_\alpha(\omega)=[e^{(\hbar\omega/k_B T_\alpha)}-1]^{-1}$ for reservoir $\alpha$ with temperature $T_\alpha$, we have 
\begin{align}
\boldsymbol{\Sigma}^<_\alpha(\omega) &=-if_\alpha(\omega)\boldsymbol{\Gamma}_\alpha(\omega),\\
\boldsymbol{\Sigma}^>_\alpha(\omega) &=-i(1+f_\alpha(\omega))\boldsymbol{\Gamma}_\alpha(\omega).
\label{greensfunc2}
\end{align}
with $\boldsymbol{\Gamma}_\alpha(\omega)=i[\boldsymbol{\Sigma}^r_\alpha(\omega)-\boldsymbol{\Sigma}^a_\alpha(\omega)]$. 
Substituting Eqs. (\ref{greensfunc1})-(\ref{greensfunc2}) into Eq. (\ref{localheatcurrent1}), we obtain a Landauer-like expression for the local heat current (see details in the Appendix \ref{appendix:Landauerform}):
\begin{equation}
    \label{jnm}
    J_{nm}=\sum_{\alpha(\neq\beta)}\int^{+\infty}_{0} \frac{d\omega}{2\pi} \hbar\omega  \mathcal{T}^{\beta\alpha}_{nm}(\omega)[f_\beta(\omega) -f_\alpha(\omega)],
\end{equation}
with
\begin{equation}
\label{tnm}
    \mathcal{T}^{\beta\alpha}_{nm}(\omega)=-2\sum_{i,j} K_{mj,ni}^C 
    \text{Im}[\boldsymbol{G}^r\boldsymbol{\Gamma}_\alpha\boldsymbol{G}^a]_{ni,mj}(\omega)\quad(\beta\neq\alpha).
\end{equation}
The current expression here is summed over all polarization directions of site $m$ and $n$.

\subsection{Local Temperature} 
\label{sec:localtemperature}
For a unique and practical definition of temperatures, an approach inspired by the zeroth law of thermodynamics has been developed \cite{stafford2013,stafford2014,stafford2015a,stafford2015b,stafford2016a,stafford2016b}, wherein, an extra reservoir coupled locally to the system is introduced as a temperature probe that the local temperature is obtained by vanishing net heat current between the probe and the system. Here we take the subscript $p$ for the temperature probe. This can be expressed as:
\begin{equation}
\label{terminal_current}
J_p=\sum_\alpha \int_0^\infty \frac{d\omega}{2\pi} \hbar\omega  \mathcal{T}^{p\alpha}(\omega)[f_p(\omega)-f_\alpha(\omega)]=0,
\end{equation}
where $\mathcal{T}^{p\alpha}(\omega)=$Tr$[\boldsymbol{G}^r\boldsymbol{\Gamma}_p\boldsymbol{G}^a\boldsymbol{\Gamma}_\alpha](\omega)$ is the transmission function between reservoir $\alpha$ and the temperature probe. Now, the self-energy term in Green's functions consists of both coupling with reservoirs $\boldsymbol{\Sigma}^r_L(\omega)+\boldsymbol{\Sigma}^r_R(\omega)$ and that with the temperature probe $\boldsymbol{\Sigma}^r_p(\omega)$. It was shown in Refs.~\cite{stafford2016a,stafford2016b} that for a wide-band and weakly coupled probe, there exists a unique solution for the local electron temperature and chemical potential. In this limit, density of states in the probe have no effect on the temperature and the local heat current is left unperturbed. 
Local temperature obtained with this method in non-equilibrium system is compatible with scanning thermal microscopic techniques, although some coarse-graining is needed due to the limited spatial resolution of the scanning probe.

To get the temperature distribution, we use the following protocol. 
In each calculation performed, the temperature probe interacts only with one atom. We use wide-band limit approximation and assume that all three vibration directions of the atom are coupled to the probe with constant strength $\gamma_p$, so there are only three non-zero diagonal elements $-i\gamma_p \omega$ in the self-energy matrix $\boldsymbol{\Sigma}^r_p(\omega)$ for the coupled atom. In Fig.~\ref{fig:flowP30detail}(c), we verified that when $\hbar\gamma_p$ is far weaker than the system energy scale (the maximum energy of phonon dispersion in pristine graphene is $\sim200$ meV), the local temperature sampled by the probe is independent of the system-probe coupling strength $\gamma_p$. The full temperature distribution is obtained by repeating the above-described process for each atom.

\section{Numerical results}
\subsection{Vortex formation due to local current injection}

Consider a graphene nanoribbon connected to two reservoirs, we use parameters $N^Z$ and $N^A$ to represent the number of hexagon cell at zigzag and armchair edge respectively. The reservoirs are connected to the longer zigzag edge through armchair leads, as shown in Fig.~\ref{fig:sketch}. The geometric parameter for armchair leads and reservoirs is given by $N^Z_{\rm Res}$. Notice that the dimension of the system is in nanometer scale, far smaller than the mean free path of phonon-phonon and phonon-electron scattering in pristine graphene at the studied temperature range \cite{bae2013,feng2015,yang2021}. In this case, ballistic transport dominates and phonons are scattered elastically. 

\begin{figure*}[!]
    \centering
    \includegraphics[scale=0.72]{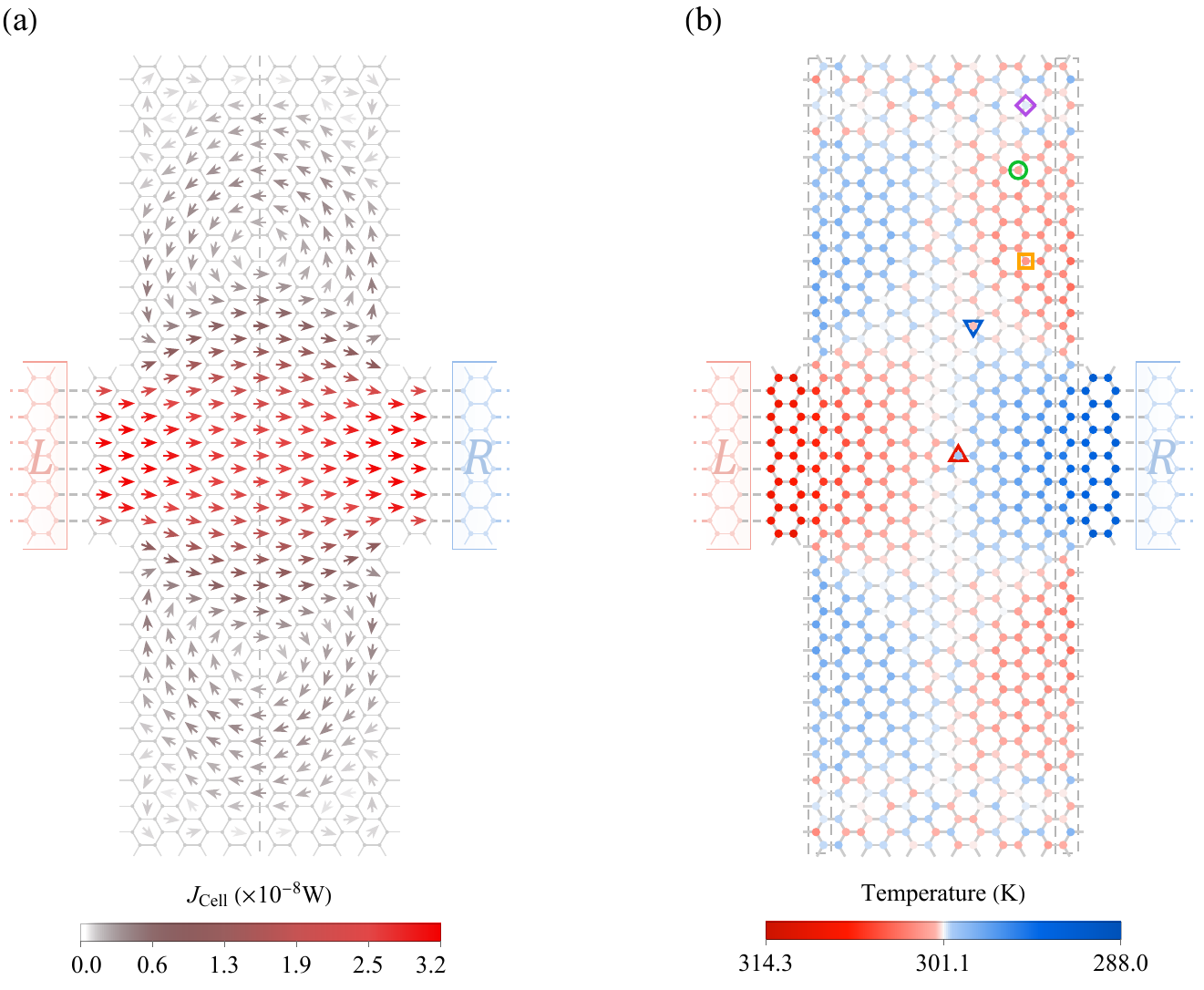}
    \caption{Spatial pattern of local heat current (a) and local temperature (b) for a temperature bias of $T_L=325$ K, $T_R=275$ K. The coupling strength of the system and the temperature probe is $\hbar\gamma_p=0.065$ meV. Geometric parameters: $N^Z=30$, $N^A=11$, and $N^Z_{\rm Res}=6$.}
    \label{fig:flowP30}
\end{figure*}

In the following numerical calculation, the spring constant matrix is generated using LAMMPS \cite{LAMMPS}, where the empirical bond order potential \cite{brenner2002} is employed for C-C and C-H interaction and Tersoff's potential \cite{tersoff1994} is used for Si-C interaction. The edges of the ribbon are passivated by hydrogen atoms. 

One example of the calculated local temperature and local heat current distribution is presented in Fig.~\ref{fig:flowP30}, with the parameters given in the figure caption. The spatial distribution of local heat current is obtained from relation $\vec{J}_{nm}=J_{nm} \Vec{e}_{nm}$ ($\Vec{e}_{nm}$ is a unit vector pointing towards site $m$ from site $n$), where $J_{nm}$ is calculated using Eqs.~(\ref{jnm}) and (\ref{tnm}). We have checked numerically the energy conservation relation $\sum_{n(\neq m)} J_{nm}=0$ in the central region. The arrows shown in Fig. \ref{fig:flowP30}(a) are given by the sum of $\vec{J}_{nm}$ among six atoms in the corresponding hexagon cell $\vec{J}_{\rm Cell}=\sum_{\{n,m\}\in \rm Cell}\vec{J}_{nm}$. Color and orientation of the arrows mark the magnitude and direction of local heat current. We can see that dominant contribution of the heat current from $L$ to $R$ comes from the region that directly connects the two leads (red region), with current vortices appearing at both sides of this region (gray region). Comparing with the local temperature distribution, we find inverted temperature distribution at vortex regions. Our results here have recovered previous results conducted in bulk graphene system \cite{shang2020,zhang2021,raya2022}. In pristine ballistic graphene nanoribbon, phonons are scattered by the borders, and heat vortices are induced by reflection on the vertical borders opposite the injection leads \cite{zhang2021}. Therefore, injected phonons are entering the temperature probe on the opposite border, resulting in inverse temperature response \cite{shytov2018}.

\begin{figure}[!]
    \centering
    \includegraphics[scale=0.62]{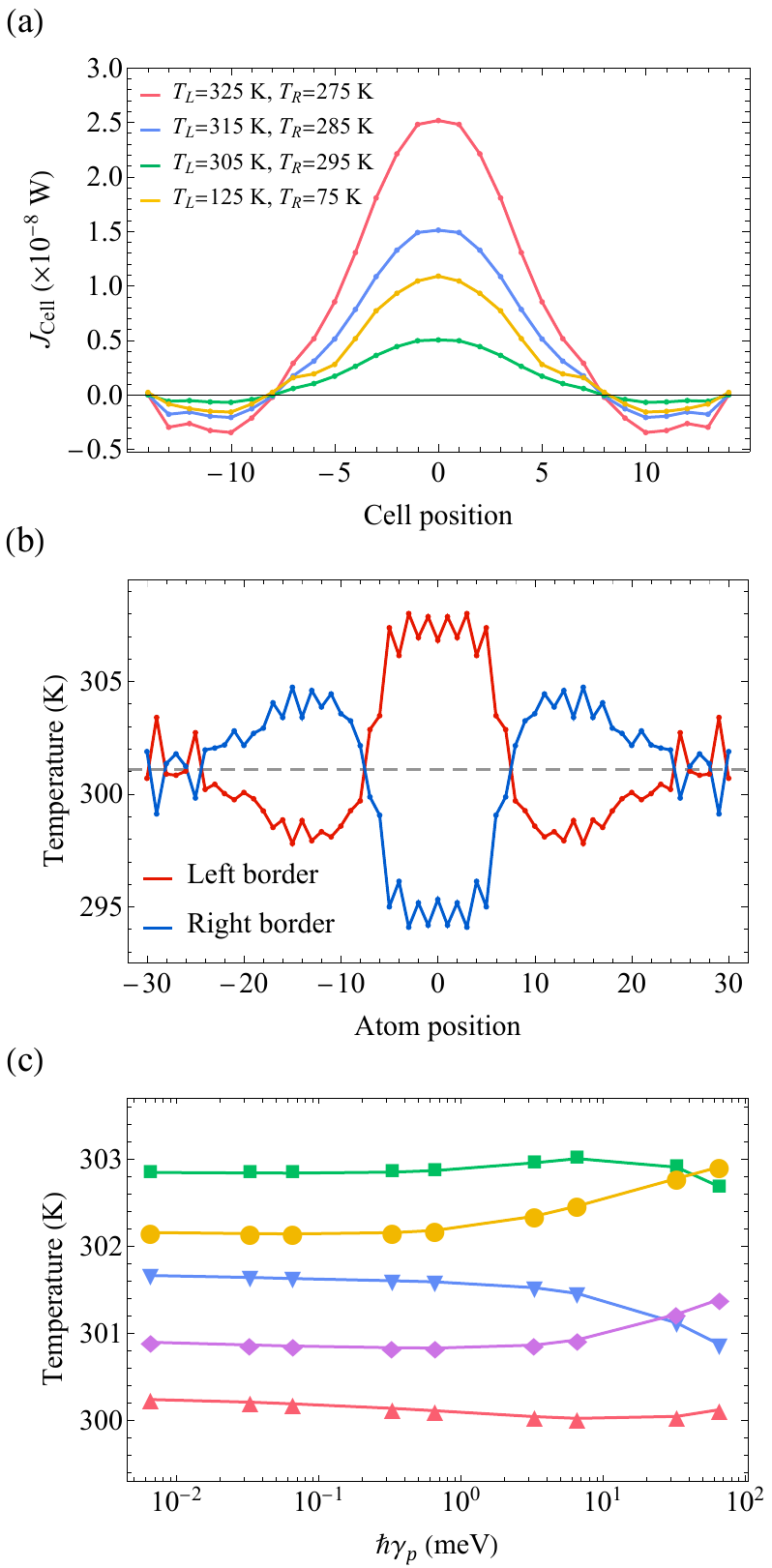}
    \caption{Detailed profile of Fig.~\ref{fig:flowP30}. (a) Local heat flux across the vertical dashed line cut in Fig.~\ref{fig:flowP30}(a). The cell position is counted from bottom to top along the line cut. 
    (b) Local temperature of the atoms on the left and right vertical border of the ribbon. The selected atoms are marked by the gray dashed boxes in Fig.~\ref{fig:flowP30}(b). The atom site in the middle is set to be position 0. Dashed line: average temperature in the ribbon.
    (c) Local temperature of the atoms shown in Fig.~\ref{fig:flowP30}(b) with same color markers as a function of the coupling parameter $\gamma_p$.}
    \label{fig:flowP30detail}
\end{figure}

Detailed thermal profiles are depicted in Fig.~\ref{fig:flowP30detail}. In Fig. \ref{fig:flowP30detail}(a), we illustrate the current flowing through the dashed line cut in Fig.~\ref{fig:flowP30}(a). The magnitude of the backflow current is approximately 10\%$\sim$15\% of the maximum mainstream in the middle, while in the bulk graphene system, it was reported to be two orders of magnitude smaller \cite{shang2020,raya2022}. The results also indicate that the temperature bias of the reservoirs has little impact on the spatial pattern of local current. In Fig.~\ref{fig:flowP30detail}(b), we extract the local temperature on the left and right border of the ribbon, marked by the dashed boxes in Fig.~\ref{fig:flowP30}(b). The average temperature in the ribbon is approximated at 301K, slightly higher than the average temperature of the reservoirs ($300$K). The inverted temperature distribution at the backflow region is quite noticeable compare to the atoms subjected to the leads in the middle. In order to verify the proclaimed assumption in Sec.~\ref{sec:localtemperature}, we check the sampled temperature by tuning the coupling strength $\gamma_p$. Results in Fig.~\ref{fig:flowP30detail}(c) confirm that, in the weak coupling limit, local temperature obtained from the probe remains unaffected by the system-probe coupling.

\begin{figure}[!]
    \centering
    \includegraphics[scale=0.62]{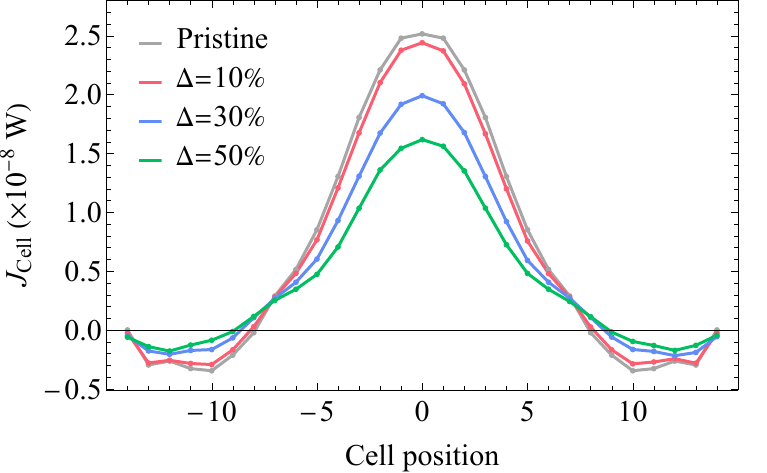}
    \caption{Local heat flux in the ribbon with mass disorder, where $\Delta$ is the disorder strength. The disordered current pattern is averaged from 50 ensembles. The rest of the parameters are the same in Fig.~\ref{fig:flowP30}.}
    \label{fig:flowP30disorder}
\vspace{8pt}

    \includegraphics[scale=0.63]{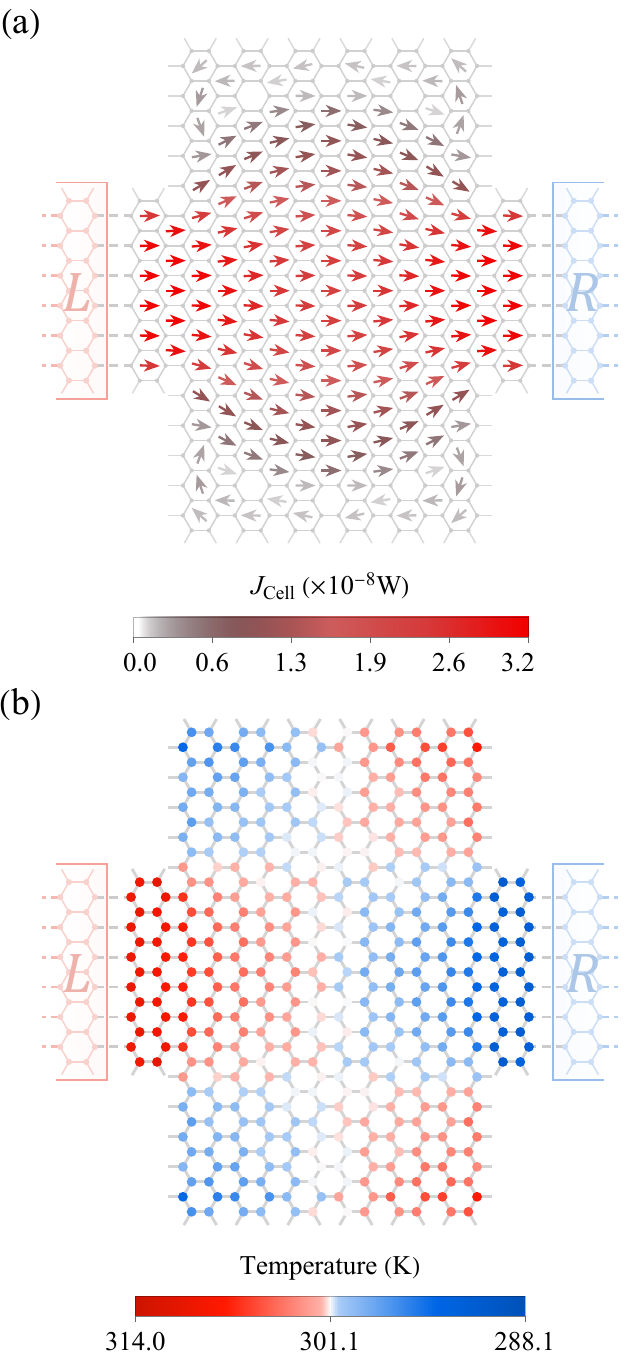}
    \caption{Thermal profile in a narrow ribbon. (a) Local heat current. (b) Local temperature. Geometric parameters: $N^Z=16$, $N^Z_{\rm Res}=6$, and $N^A=11$. The rest of the settings are the same as Fig.~\ref{fig:flowP30}.}
    \label{fig:flowP16}
\end{figure}

In a realistic setup, disturbances are inevitable, leading to inhomogeneity in the graphene nanoribbon. Here, we consider mass-disorder to model elastic scattering \cite{dhar2005,dhar2006,chaudhuri2010,lepri2016}, such global disturbance can be due to the interaction with a substrate or the presence of isotopes. Disorder in the ribbon is realized by randomly altering the mass of the atoms in spring constant $K^C$. Randomized mass $m_n'$ for the $n$-th atom is chosen from a uniform distribution between $m_n(1-\Delta)$ and $m_n(1+\Delta)$, where $\Delta$ is the disorder strength. It can be seen in Fig.~\ref{fig:flowP30disorder} that although the transmitting current decreases substantially with the increase of $\Delta$, the backflow heat current survive the disorder, even at $\Delta=50\%$. The backflow current near $\pm10$th cells indicates a slight distortion of the vortex pattern due to disorder. A similar result was mentioned by earlier study in an electron system \cite{gomes2021}.

\begin{figure}[!]
    \centering
    \includegraphics[scale=0.62]{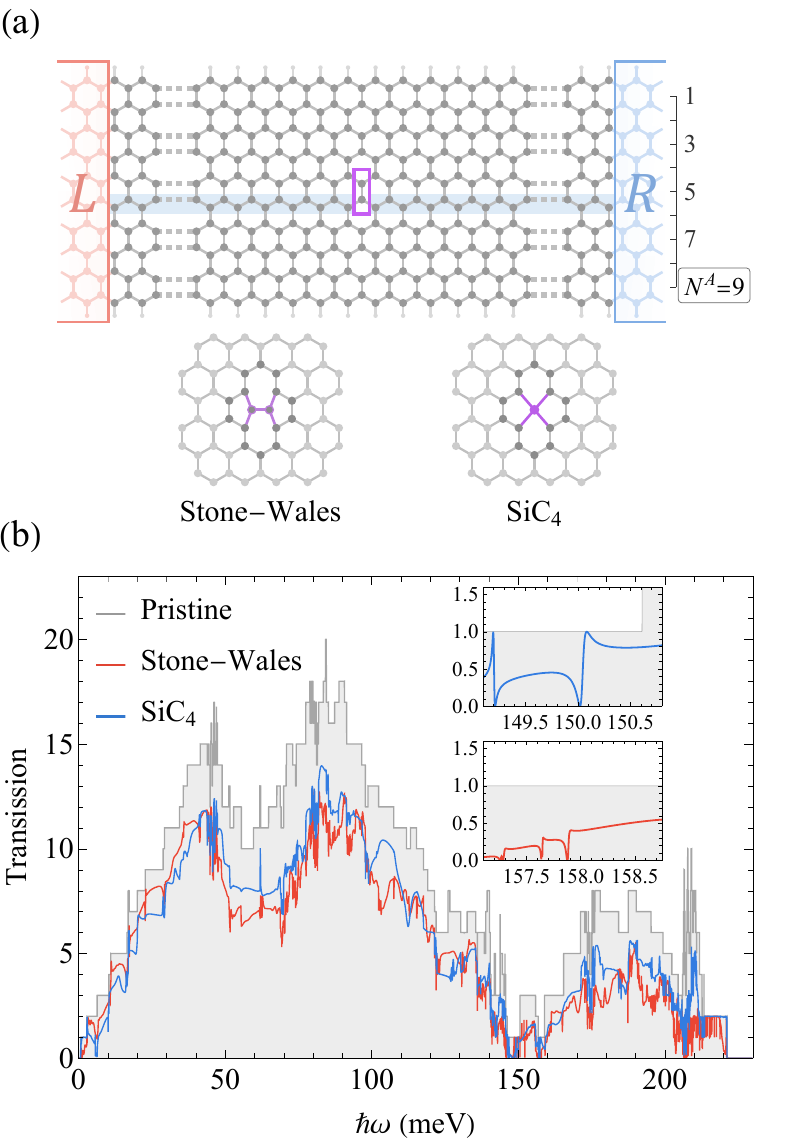}
    \caption{Zigzag graphene nanoribbon with defects. $N^A=9$ is the only parameter to mark its width. (a) A sketch of the ribbon and defects. The defect is placed at the middle marked by the purple box. (b) Transmission of zigzag graphene nanoribbon with Stone-Wales and $\mathrm{SiC_4}$ defects. Insets show phonon modes with zero transmission.}
    \label{fig:transmission}
\end{figure}

If the size of vertical borders are comparable to the horizontal ones, the current are more likely to be reflected by the horizontal borders on the top and bottom, which do not contribute to the backflow. One example with $N^Z=16$, $N^Z_{\rm Res}=6$, and $N^A=11$ is shown in Fig.~\ref{fig:flowP16}. In such a confined system, only the middle mainstream persists. The spiraling whirlpools at the corner result from competitive scattering by vertical and horizontal borders. Although the net heat current flows directly from left to right, there still exists opposite temperature response in Fig.~\ref{fig:flowP16}(b). Comparing with the local temperature distribution in Fig.~\ref{fig:flowP30}, we find that the current does not always flow from hotter to colder region.
Based on the trials conducted, we reach to the conclusion that noticeable backflow occurs when the geometric parameters satisfy a rough ratio: $N^Z-N^Z_{\rm Res}>N^A$. Furthermore, we have also checked the results in the ribbon with longer armchair edge ($N^A>N^Z$), where the leads are connected to the middle of armchair borders. These results indicate that the pattern of heat vortex is not influenced by graphene edges, while the ratio of the geometric parameters is of significance.

\begin{figure*}[ht]
    \centering
    \includegraphics[scale=0.6]{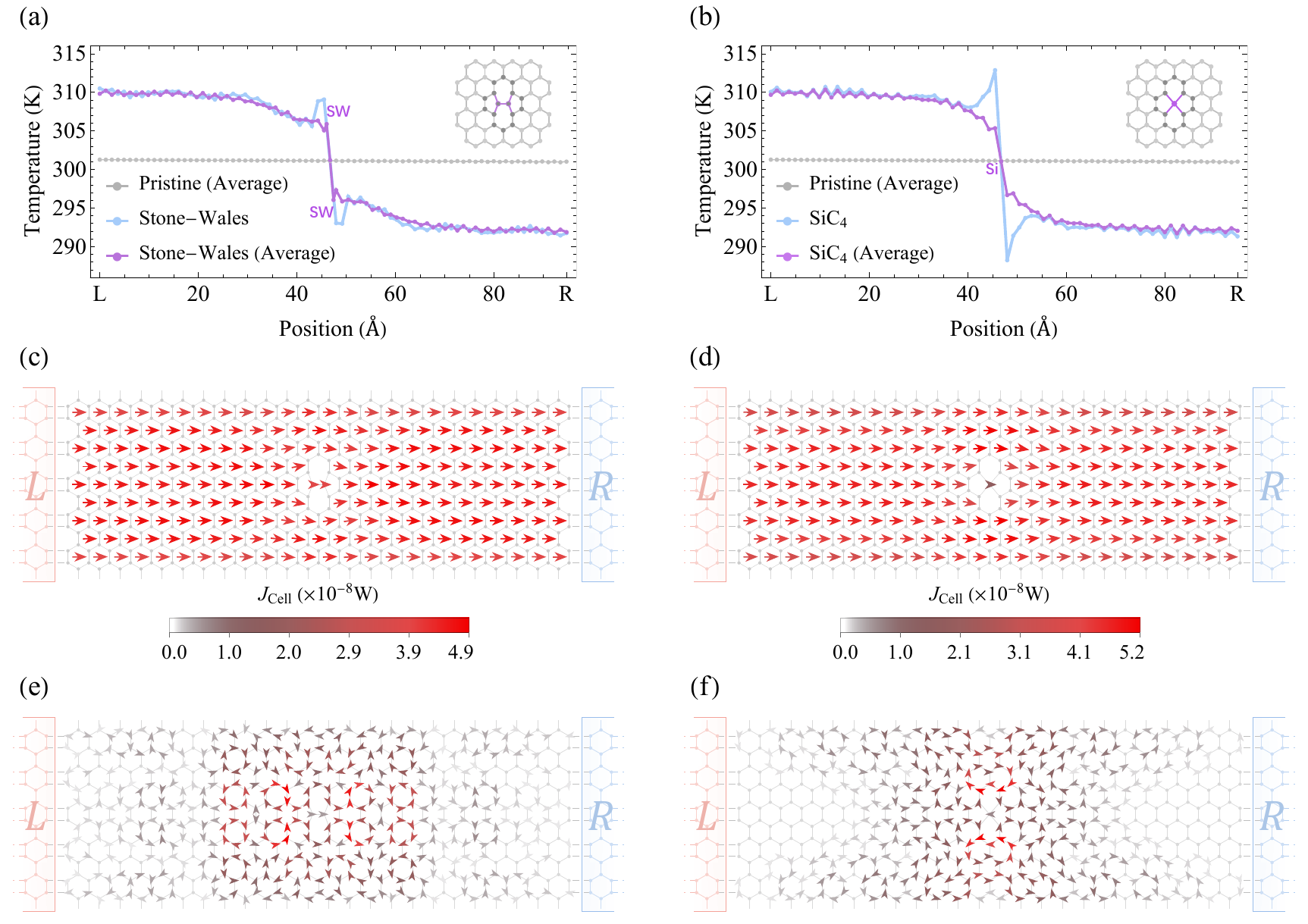}
    \caption{Local heat current and temperature of zigzag graphene nanoribbon with Stone-Wales (SW)  (a), (c), (e) and $\mathrm{SiC_4}$ (b), (d), (f) defect. Averaged local temperature in (a) and (b) is calculated by the atoms sharing the same horizontal coordinate, from the left reservoir to the right. Temperature distribution near the defect is demonstrated by light blue lines, extracted from the shaded region at the middle of the nanoribbon in Fig.~\ref{fig:transmission}(a). (c)-(d) Local heat current through the defect atoms are given by $\vec{J}_{n}=\sum_{m(\neq n)}\vec{J}_{nm}$, the color gradient is consistent with $\vec{J}_{\rm Cell}$.
    (e)-(f) Frequency resolved $\vec{J}_n$ at $\hbar\omega=157. 87$ (e) and  $150.02$ meV (f). These frequencies are found in the insets of Fig.~\ref{fig:transmission}.} 
    \label{fig:flowdefect}
\end{figure*}

\subsection{Effect of atomic defects on heat current}
\label{sec:defect}
As a source of scattering, localized defects are unavoidable in real structures. In the past decades, massive attention was paid to defect engineering aimed at tuning the functionality of atomic structures through modification of local bond configurations \cite{dattbhatt2022}. In this section, we discuss local transport features in zigzag graphene nanoribbon with localized Stone-Wales (SW) and $\mathrm{SiC_4}$ defect. A sketch of the structure is shown in the inset of Fig.~\ref{fig:transmission}(a), where the defects are placed in the middle of the ribbon. The SW defect is formed by twisting two carbon atoms by $90^\circ$ with respect to the midpoint of their bond, while the $\mathrm{SiC_4}$ defect replaces two bonded carbon atoms by a silicon atom. Here, we only need one parameter $N^A$ to describe the width of the nanoribbon since its dimension is extended parallel to the zigzag direction. 

The calculated transmission in Fig.~\ref{fig:transmission}(b) exhibits suppression to phonon transport for both defects. The insets illustrate the gap states with zero transmission near 150 meV, located between optical and acoustic branches. Figure~\ref{fig:flowdefect} shows local temperature and local heat current for SW and $\mathrm{SiC_4}$ defect. For brevity, local temperature presented in Figs.~\ref{fig:flowdefect}(a) and (b) is averaged along the transverse direction. The temperature profile manifests typical ballistic features in quasi-one-dimensional systems \cite{dhar2008,saaskilahti2012,ness2016,miao2016,nitzan2020-1}, wherein the temperature drop is induced by the defect. We notice that our result shows features of residual-resistivity dipole in electronic transport \cite{landauer1975,todorov1999,morr2017,lupke2017}, where the temperature difference at the vicinity of the defect is much larger than the overall temperature drop in the ribbon, especially around the $\mathrm{SiC_4}$ defect shown in Fig.~\ref{fig:flowdefect}(b) (light blue line). This can be attributed to the heavier silicon atom blocking heat flow in the ribbon since it only vibrates at very low frequency \cite{hage2020a,xu2023,bao2023}. As shown in Fig.~\ref{fig:flowdefect}(d), local current on the silicon atom is much weaker than those in the ribbon, where local heat current have to go bypass the defect. Therefore, extra heat is blocked in front of the defect, and depleted behind it. In contrast, the result in Fig.~\ref{fig:flowdefect}(c) shows weaker blockade by the SW defect atoms; in that case, phonons are primarily scattered by the reconstructed bonds. 

Some prior studies suggested that localized defects may give rise to vortex patterns \cite{liu2008,yamamoto2008,bao2023}, especially the gap states with zero transmission. Here, we recovered the vortex pattern for both cases, as depicted in Figs.~\ref{fig:flowdefect}(e) and (f). The specific frequencies are found at the zero transmission shown in the inset of Fig.~\ref{fig:transmission}(b). The frequency resolved heat vortices are localized near the defect, due to perfect reflection originated from quantum interference effect. However, these vortex patterns tend to be smeared out in the total heat current which includes integration from all the modes, especially those with long wavelengths. This marks one important difference between electron and phonon transport. In the former case, where particles near the Fermi level dominate the transport, current distribution depends on electron energy and the effect of quantum interference is clearly seen \cite{todorov1999,solomon2010,stegmann2020,gomes2021}, while in the latter case, phonon modes in a wide frequency range contribute, rending the interference effect more difficult to observe.

\section{Conclusions}
We studied local heat current and temperature distribution in the graphene nanoribbon using the non-equilibrium Green's function method. Inspired by the studies conducted in ballistic bulk systems, we extend the study to nanoscale graphene ribbons where wave property of phonons becomes important. We have recovered heat vortex and inverse temperature response in atomic structures, and confirmed that the boundary scattering is indeed critical in the formation of heat current vortex, whose pattern is controlled by the ratio of geometric parameters. In contrast to the results in the bulk system, localized whirlpool could appear at the corners even in a long ribbon. For further study, we expect that heat vortices can be tuned by tailoring the border geometries.
We furthermore apply our method to study Stone-Wales and $\mathrm{SiC_4}$ defects in the ribbon. In both cases, phonon transmission is reduced due to scattering by deformed bonds.
We observe heat vortices for injection of phonons at certain frequency, while these patterns disappear after including contributions from all the phonon modes. Meanwhile, we have recovered the feature of residual-resistivity dipole in the ribbon with defects. These results further extend the analogy with local electrochemical potential in electron transport.

\begin{acknowledgments}
We acknowledge financial support from the National Natural Science Foundation of China (Grant No. 22273029).
\end{acknowledgments}

\appendix
\section{Derivation of the local heat current}
\label{appendix:Landauerform}
In this Appendix, we give details of writing the local heat current using Green's function. 

In the steady state with time-translational invariance, we have $\boldsymbol{G}^<(t,t')=\boldsymbol{G}^<(t-t')$. Taking the Fourier transform of Eq.~({\ref{localheatcurrent1}}), we obtain:

\begin{align}
    \begin{split}
        J_{nm}&=\frac{1}{2}\sum_{i,j}\int^{+\infty}_{-\infty} \frac{d\omega}{2\pi} \hbar\omega
        \\
        &\quad\times\Big[K_{mj,ni}^C G^<_{ni,mj}(\omega)-K_{ni,mj}^C G^<_{mj,ni}(\omega)\Big]
    \end{split}\notag
    \\
    \begin{split}
        &=\frac{1}{2}\sum_{i,j}\int^{+\infty}_{0} \frac{d\omega}{2\pi} \hbar\omega \\
        &\quad\times\Big\{K_{mj,ni}^C \Big[G^<_{ni,mj}(\omega)-G^<_{ni,mj}(-\omega)\Big]
    \end{split}\notag
    \\
    \begin{split}
        &\qquad\quad-K_{ni,mj}^C \Big[G^<_{mj,ni}(\omega)-G^<_{mj,ni}(-\omega)\Big]\Big\}
    \end{split}\notag
    \\
    \begin{split}
        &=\frac{1}{2}\sum_{i,j}\int^{+\infty}_{0} \frac{d\omega}{2\pi} \hbar\omega \\
        &\quad\times \Big\{K_{mj,ni}^C \Big[G^<_{ni,mj}(\omega)-G^>_{mj,ni}(\omega)\Big]
    \end{split}\notag
    \\
    \begin{split}
        &\qquad\quad-K_{ni,mj}^C \Big[G^<_{mj,ni}(\omega)-G^>_{ni,mj}(\omega)\Big]\Big\},
    \end{split}
\end{align}

where relation $\boldsymbol{G}^<(-\omega)=[\boldsymbol{G}^>(\omega)]^T$ is used here. Proceeding with Eq. (\ref{greensfunc1}), we have

\begin{align}
    \begin{split}
      J_{nm}&=\frac{1}{2}\sum_{i,j}\int^{+\infty}_{0} \frac{d\omega}{2\pi} \hbar\omega \\ 
      &\quad\times\Big\{K_{mj,ni}^C [\boldsymbol{G}^r(\boldsymbol{\Sigma}^<+\boldsymbol{\Sigma}^>)\boldsymbol{G}^a]_{ni,mj}\\
      &\qquad\quad-K_{ni,mj}^C [\boldsymbol{G}^r(\boldsymbol{\Sigma}^<+\boldsymbol{\Sigma}^>)\boldsymbol{G}^a]_{mj,ni}\Big\}
    \end{split}\notag
    \\
    \begin{split}
      &=-\frac{i}{2}\sum_\alpha\sum_{i,j}\int^{+\infty}_{0} \frac{d\omega}{2\pi} \hbar\omega\\ 
      &\quad\times\Big\{K_{mj,ni}^C [\boldsymbol{G}^r\boldsymbol{\Gamma}_\alpha\boldsymbol{G}^a]_{ni,mj}\\
      &\qquad\quad-K_{ni,mj}^C [\boldsymbol{G}^r\boldsymbol{\Gamma}_\alpha\boldsymbol{G}^a]_{mj,ni}\Big\}[1+2f_\alpha(\omega)].
    \end{split}
\end{align}

One should notice that the spring constant matrix $K^{C}$  and the spectral function $i[\boldsymbol{G}^r-\boldsymbol{G^a}]=i[\boldsymbol{G}^>-\boldsymbol{G}^<]=\boldsymbol{G}^r\boldsymbol{\Gamma}\boldsymbol{G}^a$ are symmetric matrices \cite{liu2008}, yielding

\begin{equation}
    [\boldsymbol{G}^r\boldsymbol{\Gamma}\boldsymbol{G}^a](\omega)=[\boldsymbol{G}^r\boldsymbol{\Gamma}\boldsymbol{G}^a]^T(\omega),
\end{equation}
thus
\begin{equation}
    \sum_\alpha[\boldsymbol{G}^r\boldsymbol{\Gamma}_\alpha\boldsymbol{G}^a](\omega)-\sum_\alpha[\boldsymbol{G}^r\boldsymbol{\Gamma}_\alpha\boldsymbol{G}^a]^T(\omega)=0.
\end{equation}

Therefore, the expression of local heat current can be simplified:

\begin{align}
J_{nm}&=(-i)\sum_\alpha\sum_{i,j}\int^{+\infty}_{0}  \frac{d\omega}{2\pi}\hbar\omega\notag
\\
&\quad\times\Big\{K_{mj,ni}^C [\boldsymbol{G}^r\boldsymbol{\Gamma}_\alpha\boldsymbol{G}^a]_{ni,mj}(\omega)\notag
\\
&\quad\qquad-K_{ni,mj}^C [\boldsymbol{G}^r\boldsymbol{\Gamma}_\alpha\boldsymbol{G}^a]_{mj,ni}(\omega)\Big\}f_\alpha(\omega).
\end{align}

Picking one terminal, marked as $\beta$, we have 

\begin{align}
\sum_{\alpha(\neq\beta)}&[\boldsymbol{G}^r\boldsymbol{\Gamma}_\alpha\boldsymbol{G}^a](\omega)-\sum_{\alpha(\neq\beta)}[\boldsymbol{G}^r\boldsymbol{\Gamma}_\alpha\boldsymbol{G}^a]^T(\omega)\notag
\\
&\qquad=-[\boldsymbol{G}^r\boldsymbol{\Gamma}_\beta\boldsymbol{G}^a](\omega)+[\boldsymbol{G}^r\boldsymbol{\Gamma}_\beta\boldsymbol{G}^a]^T(\omega).   
\end{align}

After some algebra, we obtain Landauer-like formula for local heat current:
\begin{align}
J_{nm}&=i\sum_{\alpha(\neq\beta)}\sum_{i,j}\int^{+\infty}_{0}  \frac{d\omega}{2\pi}\hbar\omega \Big\{K_{mj,ni}^C [\boldsymbol{G}^r\boldsymbol{\Gamma}_\alpha\boldsymbol{G}^a]_{ni,mj}(\omega) \notag
\\
&\qquad-K_{ni,mj}^C [\boldsymbol{G}^r\boldsymbol{\Gamma}_\alpha\boldsymbol{G}^a]_{mj,ni}(\omega)\Big\}[f_\beta(\omega)-f_\alpha(\omega)] \notag
\\
&=-2\sum_{\alpha(\neq\beta)}\sum_{i,j}\int^{+\infty}_{0}  \frac{d\omega}{2\pi}\hbar\omega \notag
\\
&\qquad\times K_{mj,ni}^C \mathrm{Im}[\boldsymbol{G}^r\boldsymbol{\Gamma}_\alpha\boldsymbol{G}^a]_{ni,mj}(\omega)[f_\beta(\omega)-f_\alpha(\omega)].   
\end{align}

\bibliography{ref}

\end{document}